\documentstyle[aps,prb,epsf]{revtex}  

\begin{document}  

\title{Coordination defects in a-Si and a-Si:H : 
\\a characterization from first principles calculations}
\author{M. Peressi,$^{\it a}$\footnote{Corresponding author;
e-mail: {\tt peressi@ts.infn.it}} 
M. Fornari,$^{\it a}$\footnote{Present address:
Naval Research Laboratory - Code 6391, Washington DC  20375-5345}
S. de Gironcoli,$^{\it b}$ L. De Santis,$^{\it b}$
and A. Baldereschi $^{\it a,c}$}
\address{(a) Istituto Nazionale di Fisica della Materia (INFM) and \\
Dipartimento di Fisica Teorica, Universit\`a di Trieste, Strada
Costiera 11, I-34014 Trieste, Italy}
\address{ (b) Istituto Nazionale di Fisica della Materia (INFM) and  \\
Scuola Internazionale Superiore di Studi Avanzati (SISSA), via Beirut 2-4,
 I-34014 Trieste, Italy}
\address{ (c) Institut de Physique Appliqu\'ee, Ecole Polytechnique
F\'ed\'eral de Lausanne \\ PHB-Ecublens, CH-1015 Lausanne,
Switzerland}
 
\maketitle
\begin{abstract}
We study by means  of  first-principles pseudopotential method 
the coordination defects in a-Si  and a-Si:H,
also in their formation and their evolution upon hydrogen interaction.
An accurate analysis of the valence charge distribution
and of the ``electron localization function'' (ELF) allows to resolve
possible ambiguities in the bonding configuration, and  in particular
to identify clearly   three-fold  ($T_3$)  and five-fold  ($T_5$) 
coordinated defects.
We found that electronic states in the gap can be associated
to both kind of defects,  and that in both cases
the interaction with hydrogen can reduce 
the density of states in the gap.
\end{abstract}
~
\section{Introduction}
Amorphous silicon (a-Si) and hydrogenated amorphous silicon (a-Si:H)
are  prototypes of disordered covalent semiconductors.
Extensive work, both experimental and theoretical, has been done to study
their topological and electronic structure.
Although most of Si atoms are tetrahedrally coordinated, anomalously 
coordinated configurations can locally occur in pure and hydrogenated 
amorphous samples, but 
--at variance with the case of crystals where coordination 
defects can be easily recognized as deviations from the perfect 
 ordered structure-- their identification  is not trivial.
Hence, one of the most challenging problems in the amorphous systems
is to localize the defects, to classify them and to identify their 
peculiar electronic features.

Traditionally, three-fold ($T_3$) defects
have been considered as the most likely intrinsic defects  in a-Si.
The non vanishing  density of states (DOS)  
observed in the gap has been commonly ascribed  for a long time to the
 ``dangling bonds'' corresponding to these defects,
and its lowering upon hydrogenation has been explained with the 
saturation of dangling bonds by hydrogen 
(Ley 1984, Fedders and Carlsson 1988, 1989,
Biswas {\it et al.} 1989, Holender and Morgan 1993,
Lee and Chang 1994, Davis 1996, Tuttle and Adams 1996).

More recently, this picture has been debated and revised.
In particular the
importance of five-fold coordinated ($T_5$ or  ``floating bonds'') in a-Si
has been clearly stated in the theoretical works by 
Pantelides (1986, 1987) and 
Kelires and Tersoff (1988) a douzen of years ago,
both in terms of their {\it existence} and  their peculiar
role in the electronic structure.
The  empirical simulation by 
Kelires and Tersoff (1988) has shown that $T_5$
atoms have lower energy than $T_3$ atoms, and therefore should be favoured 
in general. Also some {\it ab-initio}
molecular dynamics simulations of a-Si structures show a predominance 
of $T_5$ defects with respect to $T_3$ (Buda {\it et al.} 1989,
$\rm \check S$tich {\it et al.} 1991, Buda {\it et al.} 1991). 
Pantelides (1986, 1987) 
argued that $T_3$ and $T_5$ are {\it conjugated defects}
and must be considered on the same footing,
since a bond elongation can transform a $T_5 + T_4$
structure into a $T_4+T_3$ one, or vice versa an inward
relaxation can transform a $T_4+T_3$ structure
into a $T_5 + T_4$ one;
furthermore, he proposed a mechanism for H diffusion based on
floating-bond switching and  annihilation/formation of $T_5$'s through 
interaction with H (Pantelides 1987), which 
---at variance with the commonly accepted picture of dangling bonds
hydrogenation--- is compatible 
with the rapid decrease in the number of defects 
without any appreciable change in the density of Si--H bonds
experimentally observed at low temperature.

Some of these ideas have been widely used in discussing the 
{\it geometrical} characterization of defects;  their soundness
in terms of electronic properties has  been investigated 
 mainly   by model  calculations (Fedders and Carlsson 1987, 1988, 1989,
Fedders {\it et al.} 1992)
and more recently  by some
first-principles calculations (Fedders {\it et al.} 1992, Lee and Chang 1994,
Tuttle and Adams 1996, 1998, 
Fornari  {\it et al.} 1999).

It remains 
the necessity of a simple tool going beyond purely geometrical criteria
for a localization and an unambiguous characterization of defects.
Recently the maximally-localized Wannier function approach 
has been applied to analyze the bonding properties in
amorphous silicon ((Marzari and Vanderbilt 1997,
Silvestrelli {\it et al.} 1998).
We focus in the present work on a real-space 
analysis of the bonding pattern of a-Si and a-Si:H using the simplest
tools provided by first-principles electronic structure calculations:
a comparative analysis of 
the electronic charge density and the ``electron localization function'' 
(ELF) (Savin {\it et al.} 1992).
We  address the reader to another work (Fornari {\it et al.} 1999) for 
an accurate analysis using  local or 
projected density of states (DOS) which completes the characterization
of the coordination defects in terms of electronic properties, 
and we recall here only the main results.

\section{Results and discussions}
For studying the bonding properties in a-Si and a-Si:H
we start from some selected samples generated by other
authors (Buda {\it et al.} 1989,
$\rm \check S$tich {\it et al.} 1991, Buda {\it et al.} 1991)
 using Car-Parrinello 
first-principles molecular dynamics (CPMD).  
These structures reproduce quite well the
experimental pair correlation function and bond angle distribution
function using a reasonable number of atoms and hence they are
suitable for accurate ab-initio studies.
The configurations studied are 
cubic supercells of side $a=2\ a_0$, where $a_0$=10.17 a.u. is the
theoretical  equilibrium lattice parameter of c-Si, 
which also corresponds ---in our calculations--- to the optimized
density of a-Si and a-Si:H. The supercells contain
respectively 64 Si atoms to describe a-Si 
(Buda {\it et al.} 1989,
$\rm \check S$tich {\it et al.} 1991)
and 64 Si atoms plus 8 H atoms for a-Si:H
(Buda {\it et al.} 1989, 1991).

We use state-of-the-art electronic structure methods based on DFT using
norm-conserving pseudopotentials and plane-wave basis set (Fornari {\it et
al.} 1999).
The CPMD configurations, aiming mainly at reproducing the structural 
properties, have been obtained using a kinetic energy cutoff $E_{cut}$=12 Ry
and the $\Gamma$ point only for Brillouin Zone (BZ) sampling.
We improve  in our calculations the BZ sampling using
4 inequivalent special {\bf k} points for self-consistency and 
75  {\bf k} points for DOS.
These parameters have been chosen as
a reasonable compromise between accuracy and computational cost.
The optimization of the a-Si and a-Si:H structures with the new computational
parameters is accompanied only by small structural rearrangements, and
therefore the mean structural properties are very
similar to those reported by Buda {\it et al.} (1989, 1991) and
$\rm \check S$tich {\it et al.} (1991)
for the original configurations, and we do not discuss them in detail here.
We only report that in a-Si the mean bond length is $d
\simeq 4.47$ a.u., quite similar to the
crystalline one which is 4.40 a.u.. 
The mean bond angle is  $\vartheta\simeq 109^{\circ}$, close to the 
characteristic value of the perfect tetrahedral network. 
The location of the first minimum of the radial distribution function 
defines geometrically the cutoff distance for the nearest neighbours (NN), 
which turns out to be  $R_{NN} = 5.08$ a.u., giving an average
coordination number of about $4.03$.  
In a-Si:H  the average Si-Si bond length is 
the same as in a-Si, but the first peak of the radial distribution function is
more broadened and it is more appropriate to consider a larger NN
cutoff distance,
 $R_{NN} = 5.49$ a.u.. Each H is bound to one Si
atom with an average distance $d_H =$ 2.95 a.u., very close to the
corresponding value in SiH$_4$ molecule.

The standard {\it geometrical} analysis based simply on counting 
the atoms lying inside a sphere of radius $R_{NN}$ indicates  that
the starting  configurations have a predominance of $T_5$ defects and of
distorted $T_4$ sites. Moreover, the a-Si samples do not contain
well defined $T_3$ defects.
This feature can be a consequence of the rapid quench from the liquid 
states which has been done in preparing the sample in the molecular dynamics
process (since the liquid state is sixfold coordinated, a rapid quench
typically favours overcoordination rather than undercoordination).

We thus start analyzing in detail an overcoordinated environment.
For the sake of clarity, we will consider the case of a-Si
(in a-Si:H overcoordination can  be due to five Si neighbours, or to
four Si and one H, and so on).

\begin{figure}%%%%%%%%%%% fig 1
\hspace{1cm}\epsfbox{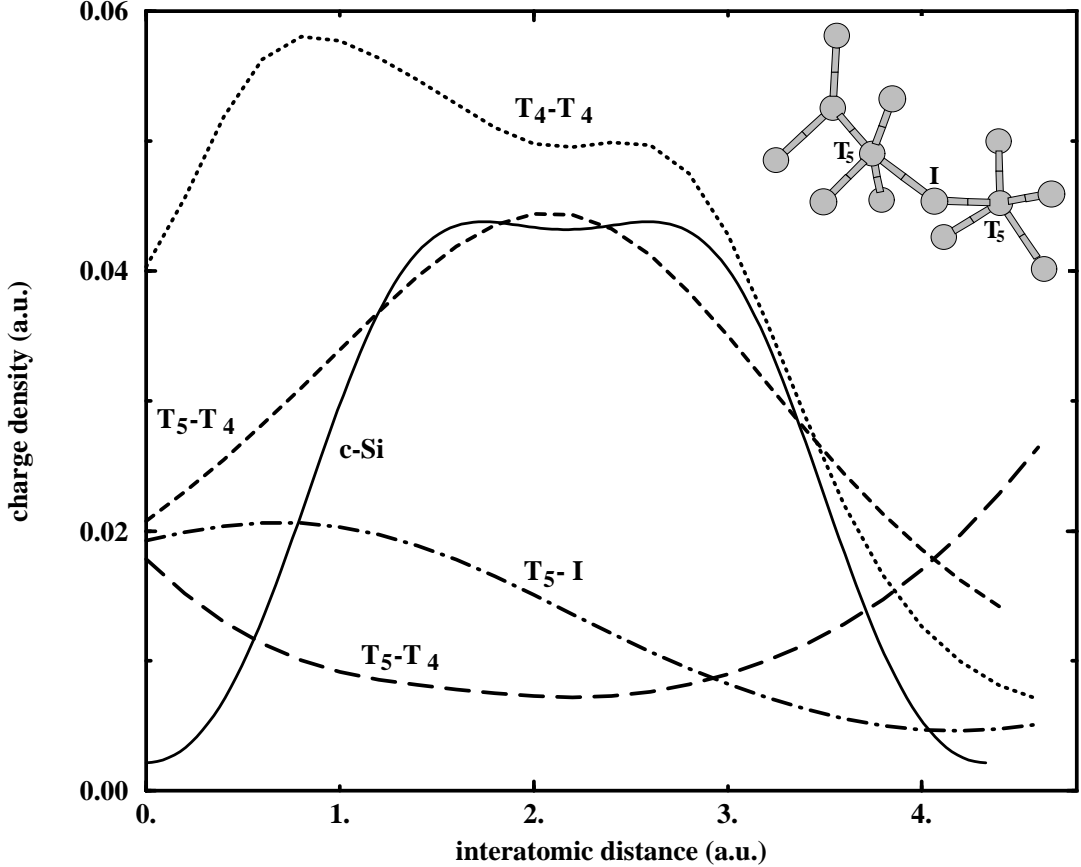}
\caption{The inset shows a snapshot from the a-Si configuration
including two $T_5$ defects with an intermediate  four-fold coordinated
atom (I) and other $T_4$ atoms (not individually labelled).
The curves are the profile of the valence 
charge density along some bonds shown in the snapshot, connecting:
two normally four-fold coordinated atoms (dotted line), 
a  $T_5$ with a  $T_4$ at ``normal'' distance (short-dashed line), 
a  $T_5$ with a  $T_4$ at ``longer'' distance (long-dashed line), 
a  $T_5$ with $I$ (dashed-dotted line). 
``Normal'' and ``longer'' are with respect to the average bond length. 
For comparison also the perfect crystalline bond is shown (solid line).
In order to filter out possible unrelevant local fluctuations,
the  charge
distribution is filtered averaging over small spheres of radius $R =$0.6 a.u.
moving along the bond.
The distance is along the geometrical bond, calculated from one 
of the two atoms connected; note the different bond lengths.
At variance with the crystalline case, the charge profiles are asymmetric
with respect to the bond centre, indicating a partially ionic character
of the bond.}
\label{fig1} 
\end{figure}

In our a-Si sample there are two $T_5$
sites  close one to each other (labelled A and B in the upper snapshot in
figure 1), with a
sort of interstitial (I) atom connecting them.
A charge density analysis confirms for this configuration
the bonding pattern predicted by 
 the {\it geometrical} criteria, and 
helps in characterizing the different types of
bonds   ($\rm \check S$tich {\it et al.} 1991). 
We observe that $T_5$ sites are accompanied by a 
valence charge density depletion.
The charge density profiles reported in 
figure 1 show in particular that  some $T_5$-$T_4$ ``long'' bonds
 and the bonds $T_5$-$I$ are characterized by a very small
charge density;  hence, they  are ``weak'' and
therefore those $T_5$ defects are the best candidates to transform into
$T_3$ sites after a bond elongation.
The asymmetry in the bond charge profiles 
indicates that, at variance with the perfect crystalline
environment, the bonds are not perfectly homopolar but have a certain 
degree of ionicity.

It is useful to investigate the bonding pattern using a different kind of
real-space analysis, i.e. the study of the ``electron localization function''.
The ELF was originally defined as a scalar function ${\cal E}({\bf r})$ 
measuring the conditional
probability of finding an electron in the neighbourhood of another electron
with the same spin.  
In the reformulation due to Savin et al. (1992)  it is expressed as:
$${\cal E}({\bf r})={1\over
1+[D({\bf r})/D_h({\bf r})]^2},$$
where  $D({\bf r})$ is
 the Pauli excess energy density, i.e.
the difference between the kinetic energy density of the system and the kinetic
energy of a non-interacting system of bosons at the same density.
$D_h({\bf r})$ is the same quantity for the homogeneous electron gas at a
density equal to the local density. With this definition, a value of
${\cal E}({\bf r})$  close to 0.5 in the bonding regions
indicates a metallic character; a value close to one is characteristic of
regions  where the electrons are paired to form a covalent bond, 
but also of regions with an unpaired lone electron localized, thus
corresponding to a dangling bond.
The ELF has been originally proposed in the
 all-electron formalism, and only very recently it has been successfully
applied
in the framework of the density functional theory (DFT) within the
pseudopotential method  (De Santis and Resta 1999). 
Whereas  charge density plots are a standard tool in the first-principles
theoretical studies of real materials, ELF investigations are still lacking,
and this is, to our knowledge, the first application to disordered
solid state systems.

In the case of normal or floating bonds, the ELF does not add
much more informations with respect to the standard charge density analysis.
In the left upper panel of figure 2 we show the ELF=0.85 isosurfaces 
for the overcoordinated environment in a-Si described before.
High-value charge density (not show here) and ELF 
isosurfaces are almost similar in their extension and shape.
The  ELF isosurface in correspondence
to the A--I bond clearly visualizes its bowing (the isosurface
is not perfectly centred on the geometrical bond) and its weakness
(the isosurface is smaller than those on the other bonds). 

\begin{figure}%%%%%%%%%%% fig 2
\epsfbox{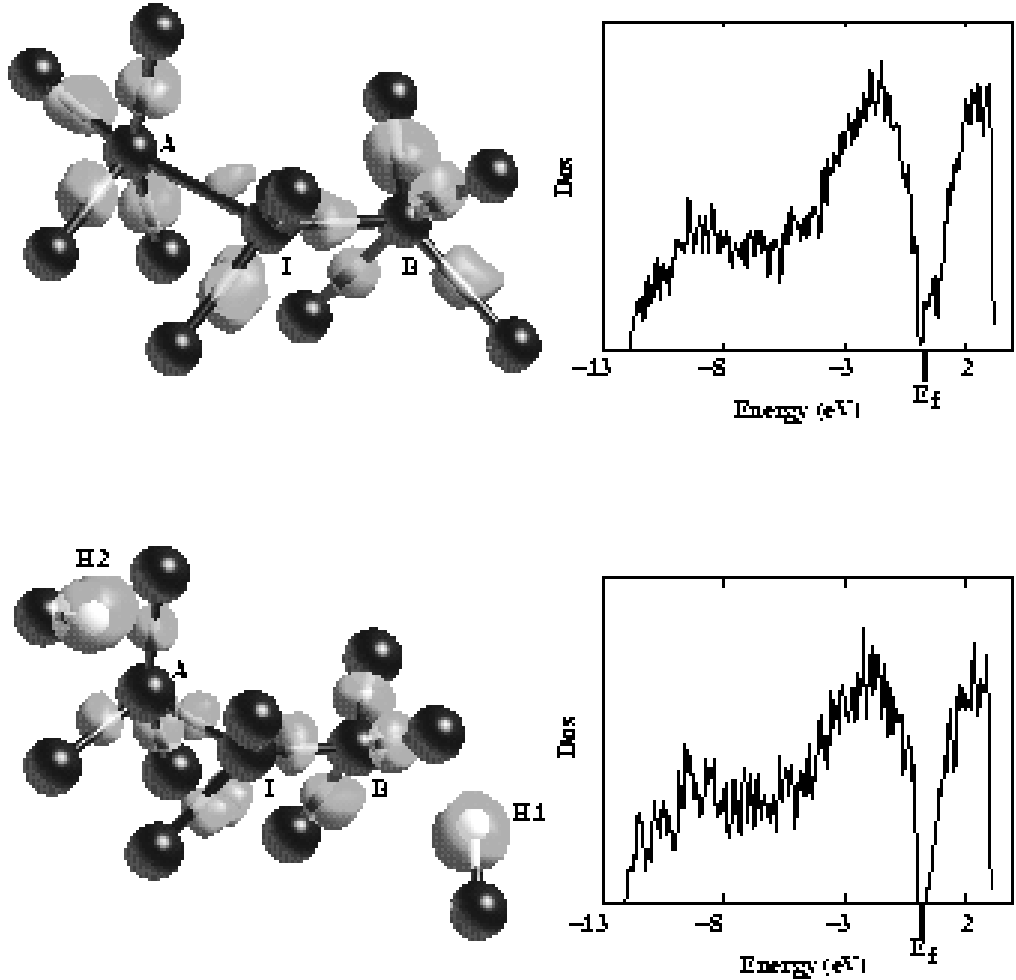}
\caption{Upper panels: a snapshot from the starting a-Si configuration
including two $T_5$ defects (A and B) with an intermediate $T_4$ atom (I).
The ELF=0.85 isosurfaces are plotted: it is clearly visible
the off-centered position
and the small distorted shape  between atoms A and I,
indicating the bowing and the weakness of the bond.
The total DOS of the a-Si sample is shown on the right, with electronic  states
close to the Fermi energy $E_f$. 
Lower panels: the structure evolved after addition of two H atoms 
(small white balls, H1 and H2) which 
have annihilated the $T_5$ defects. All the Si atoms of our sample 
are now ``normal''  $T_4$ sites, 
as shown by the more regular shape of the ELF
isosurfaces and by the vanishing DOS in the gap.}
\label{fig2} 
\end{figure}

Adding two hydrogen atoms  in the neigbourhood of the
$T_5$ sites and allowing the system to relax,
two Si--Si bonds are broken so that the atoms A and B become normally
tetrahedrally coordinated, and their fifth NN atoms connect with
the additional hydrogens (see the snapshot in the lower panel
of figure 2). In this configuration all the Si--Si bonds are rather
strong (the ELF isosurface between A and I is more extended with
respect to the previous case) and 
more bulk-like (all the isosurfaces are more regular in shape).
The plots of the density of states (right panels in figure 
2) show that,
at variance with the starting configuration having a metallic character 
evidently due to defect induced states in the gap, the final one is clearly
semiconducting.  

\begin{figure}%%%%%%%%%%% fig 3
\epsfbox{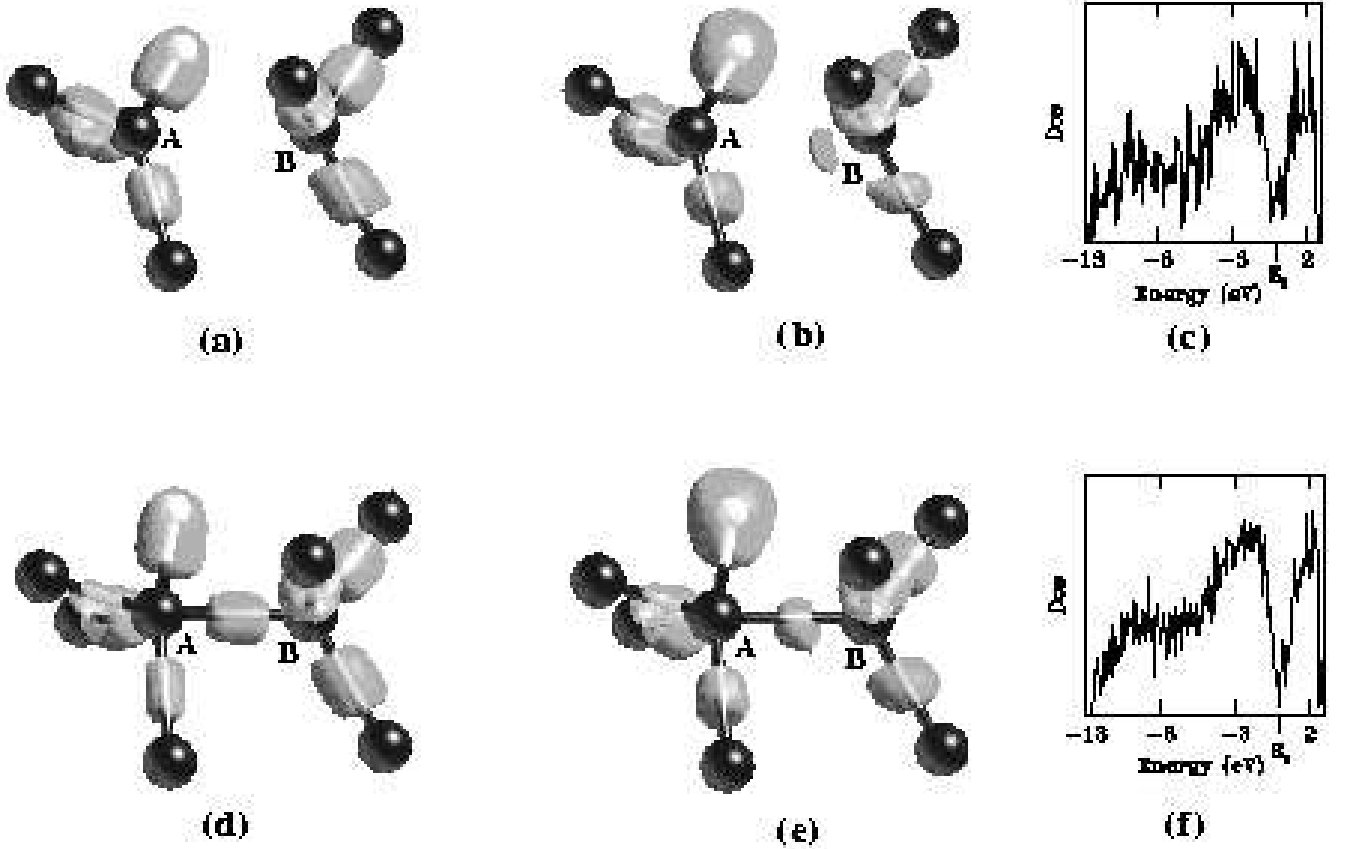}
\caption{Upper panels: a snapshot from a-Si:H with a $T_4+T_3$ 
structure (atom A and  B respectively). 
Charge density isosurfaces ($n$=0.06 a.u.) and
ELF=0.85 isosurfaces are plotted respectively in panels (a) and (b).
The dangling bond on atom B is evident by comparison of the two plots
as a region of low charge density but high ELF. 
The total DOS of the sample is reported in panel (c):
the sizeable DOS around $E_f$ is due to the $T_3$ defect
and other defects in the sample.\\
Lower panels: a snapshot containing the same atoms after relaxation, 
with a new bond formed between A and B atoms giving rise to a $T_5+T_4$ 
structure. Charge density and ELF isosurfaces are plotted in 
panels (d) and (e) respectively, as in the previous case. The new A--B bond 
is characterized as a region of high charge density and ELF.
The total DOS of the sample is reported in panel (f): gap states 
are still present.
}
\label{fig3}
\end{figure}

The combined charge-density and ELF analysis is necessary  to identify 
unambiguously the dangling bonds and to distinguish for instance a $T_5+T_4$
configuration from a $T_4+T_3$.
 Whereas the presence of a covalent 
bond is indicated by a region of local maxima of both ELF and charge density,
a dangling bond is identified by a region with high values of ELF 
but low electronic charge density.
This is evident in figure 3 (upper panels), 
where we show a snapshot from
a a-Si:H sample with a $T_4$ (labelled A) and a $T_3$ (labelled B) atoms
(we have created  a dangling bond by removing an hydrogen
 initially  bond to the silicon atom B).
Panel (a) shows charge density isosurfaces, and
panel (b)  ELF isosurfaces. 
The absence of high-value charge density isosurface {\it together with}
the presence of  high-value ELF isosurfaces in the region between atoms A and
B clearly indicate the presence of a dangling bond originated from atom B.
As expected, 
this configurations has a metallic character, with electronic states
around the Fermi energy $E_f$ (panel (c) of figure 3).

When the system is allowed to relax, a new bond is formed between 
the silicon atoms A and B, as it clear from the panels (d) (charge density)
and (e) (ELF). The final system has still gap states, both because
of the $T_5$ defect B which is now formed and because other coordination
defects are present in the rest of the a-Si:H sample.
The evolution of this structure from a  
$T_4+T_3$  into a $T_5+T_4$ is consistent with the picture of Pantelides
(1986) of the conjugated $T_3$ and $T_5$ sites. 

\section{Summary}
In conclusion, we 
have presented the results of  accurate {\it ab initio} self-consistent
pseudopotential calculations of a-Si and a-Si:H samples with
different coordination  defects 
starting from some configurations generated via
CPMD, and we have followed 
some possible processes of defect formation, annihilation by H, and
transformation of one defect into the other.
Those techniques allowing to identify the defects in real space
are suitable for their localization in disordered structures.
In particular, we have shown that 
a combined analysis of the electronic charge density distribution
and ELF allows to unambiguously classify the different kind of defects.
We have clearly identified $T_3$ and $T_5$ defects, and comparing 
the DOS in the different configurations we have shown that they 
both can induce  states in the gap, whose density 
is reduced in both cases by interaction with H.

\section{Acknowledgments}
This work has been done within the ``Iniziativa Trasversale di Calcolo
Parallelo'' of INFM. We acknowledge useful discussions with N. Marzari.
One of the authors (S. de G.) acknowledges support from the MURST
within the initiative {\it Progetti di ricerca di rilevante interesse
nazionale}.
%%%%%%%%%%%%%%%%%% Bibliografia !!!!
%%%\newpage
\setlength{\baselineskip}{0.5cm}

%%%%%%%%%%%%%%%%%%%%%%%%%%%%%%%%%%%%%%%%%%%%%%%%%%%%%%%%%%%%%%%%%%%%%%%%%%%%%%%
\end{document}